%
%
%
\def\unredoffs{} \def\redoffs{\voffset=-.31truein\hoffset=-.59truein}
\def\speclscape{\special{ps: landscape}}
%
%
%
%
%
\newbox\leftpage \newdimen\fullhsize \newdimen\hstitle \newdimen\hsbody
\tolerance=1000\hfuzz=2pt
\catcode`\@=11 
%
\ifx\answ\bigans\message{(This will come out unreduced.}
\magnification=1200\unredoffs\baselineskip=16pt plus 2pt minus 1pt
\hsbody=\hsize \hstitle=\hsize 
\else\message{(This will be reduced.} \let\l@r=L
\redoffs \hstitle=8truein\hsbody=4.75truein\fullhsize=10truein\hsize=\hsbody
\output={\ifnum\pageno=0 
  \shipout\vbox{\speclscape{\hsize\fullhsize\makeheadline}
    \hbox to \fullhsize{\hfill\pagebody\hfill}}\advancepageno
  \else
  \almostshipout{\leftline{\vbox{\pagebody\makefootline}}}\advancepageno
  \fi}
\def\almostshipout#1{\if L\l@r \count1=1 \message{[\the\count0.\the\count1]}
      \global\setbox\leftpage=#1 \global\let\l@r=R
 \else \count1=2
  \shipout\vbox{\speclscape{\hsize\fullhsize\makeheadline}
      \hbox to\fullhsize{\box\leftpage\hfil#1}}  \global\let\l@r=L\fi}
\fi
%
\newcount\yearltd\yearltd=\year\advance\yearltd by -1900

\def\Title#1#2{\nopagenumbers\abstractfont\hsize=\hstitle\rightline{#1}%
\vskip 1in\centerline{\titlefont #2}\abstractfont\vskip .5in\pageno=0}
%
%

\def\draftmode{\message{ DRAFTMODE }\def\draftdate{{\rm preliminary draft:
\number\month/\number\day/\number\yearltd\ \ \hourmin}}%
\headline={\hfil\draftdate}\writelabels\baselineskip=20pt plus 2pt minus 2pt
 {\count255=\time\divide\count255 by 60 \xdef\hourmin{\number\count255}
  \multiply\count255 by-60\advance\count255 by\time
  \xdef\hourmin{\hourmin:\ifnum\count255<10 0\fi\the\count255}}}
\def\nolabels{\def\wrlabeL##1{}\def\eqlabeL##1{}\def\reflabeL##1{}}
\def\writelabels{\def\wrlabeL##1{\leavevmode\vadjust{\rlap{\smash%
{\line{{\escapechar=` \hfill\rlap{\sevenrm\hskip.03in\string##1}}}}}}}%
\def\eqlabeL##1{{\escapechar-1\rlap{\sevenrm\hskip.05in\string##1}}}%
\def\reflabeL##1{\noexpand\llap{\noexpand\sevenrm\string\string\string##1}}}
\nolabels
%
\global\newcount\secno \global\secno=0
\global\newcount\meqno \global\meqno=1
\def\newsec#1{\global\advance\secno by1\message{(\the\secno. #1)}
\global\subsecno=0\eqnres@t\noindent{\bf\the\secno. #1}
\writetoca{{\secsym} {#1}}\par\nobreak\medskip\nobreak}
\def\eqnres@t{\xdef\secsym{\the\secno.}\global\meqno=1\bigbreak\bigskip}
\def\sequentialequations{\def\eqnres@t{\bigbreak}}\xdef\secsym{}
\global\newcount\subsecno \global\subsecno=0
\def\subsec#1{\global\advance\subsecno by1\message{(\secsym\the\subsecno. #1)}
\ifnum\lastpenalty>9000\else\bigbreak\fi
\noindent{\bf\secsym\the\subsecno. #1}\writetoca{\string\quad
{\secsym\the\subsecno.} {#1}}\par\nobreak\medskip\nobreak}

\global\newcount\subsubsecno \global\subsubsecno=0
\def\subsubsec#1{\global\advance\subsubsecno
by1\message{(\secsym\the\subsecno.\the\subsubsecno. #1)}
\ifnum\lastpenalty>9000\else\bigbreak\fi
\noindent{\bf\secsym\the\subsecno.\the\subsubsecno. #1}\writetoca{\string\quad
{\secsym\the\subsecno.\the\subsubsecno.} {#1}}\par\nobreak\medskip\nobreak}
\def\appendix#1#2{\global\meqno=1\global\subsecno=0\xdef\secsym{\hbox{#1.}}
\bigbreak\bigskip\noindent{\bf Appendix #1. #2}\message{(#1. #2)}
\writetoca{Appendix {#1.} {#2}}\par\nobreak\medskip\nobreak}
%
%
\def\eqnn#1{\xdef #1{(\secsym\the\meqno)}\writedef{#1\leftbracket#1}%
\global\advance\meqno by1\wrlabeL#1}
\def\eqna#1{\xdef #1##1{\hbox{$(\secsym\the\meqno##1)$}}
\writedef{#1\numbersign1\leftbracket#1{\numbersign1}}%
\global\advance\meqno by1\wrlabeL{#1$\{\}$}}
\def\eqn#1#2{\xdef #1{(\secsym\the\meqno)}\writedef{#1\leftbracket#1}%
\global\advance\meqno by1$$#2\eqno#1\eqlabeL#1$$}
%
\newskip\footskip\footskip14pt plus 1pt minus 1pt 
\def\footnotefont{\ninepoint}\def\f@t#1{\footnotefont #1\@foot}
\def\f@@t{\baselineskip\footskip\bgroup\footnotefont\aftergroup\@foot\let\next}
\setbox\strutbox=\hbox{\vrule height9.5pt depth4.5pt width0pt}
\global\newcount\ftno \global\ftno=0
\def\foot{\global\advance\ftno by1\footnote{$^{\the\ftno}$}}
%
\newwrite\ftfile
\def\footend{\def\foot{\global\advance\ftno by1\chardef\wfile=\ftfile
$^{\the\ftno}$\ifnum\ftno=1\immediate\openout\ftfile=foots.tmp\fi%
\immediate\write\ftfile{\noexpand\smallskip%
\noexpand\item{f\the\ftno:\ }\pctsign}\findarg}%
\def\footatend{\vfill\eject\immediate\closeout\ftfile{\parindent=20pt
\centerline{\bf Footnotes}\nobreak\bigskip\input foots.tmp }}}
\def\footatend{}
%
%
\global\newcount\refno \global\refno=1
\newwrite\rfile
\def\ref{[\the\refno]\nref}
\def\nref#1{\xdef#1{[\the\refno]}\writedef{#1\leftbracket#1}%
\ifnum\refno=1\immediate\openout\rfile=refs.tmp\fi
\global\advance\refno by1\chardef\wfile=\rfile\immediate
\write\rfile{\noexpand\item{#1\ }\reflabeL{#1\hskip.31in}\pctsign}\findarg}
\def\findarg#1#{\begingroup\obeylines\newlinechar=`\^^M\pass@rg}
{\obeylines\gdef\pass@rg#1{\writ@line\relax #1^^M\hbox{}^^M}%
\gdef\writ@line#1^^M{\expandafter\toks0\expandafter{\striprel@x #1}%
\edef\next{\the\toks0}\ifx\next\em@rk\let\next=\endgroup\else\ifx\next\empty%
\else\immediate\write\wfile{\the\toks0}\fi\let\next=\writ@line\fi\next\relax}}
\def\striprel@x#1{} \def\em@rk{\hbox{}}
\def\lref{\begingroup\obeylines\lr@f}
\def\lr@f#1#2{\gdef#1{\ref#1{#2}}\endgroup\unskip}

\def\addref#1{\immediate\write\rfile{\noexpand\item{}#1}} 
\def\startrefs#1{\immediate\openout\rfile=refs.tmp\refno=#1}
\def\xref{\expandafter\xr@f}\def\xr@f[#1]{#1}
\def\refs#1{\count255=1[\r@fs #1{\hbox{}}]}
\def\r@fs#1{\ifx\und@fined#1\message{reflabel \string#1 is undefined.}%
\nref#1{need to supply reference \string#1.}\fi%
\vphantom{\hphantom{#1}}\edef\next{#1}\ifx\next\em@rk\def\next{}%
\else\ifx\next#1\ifodd\count255\relax\xref#1\count255=0\fi%
\else#1\count255=1\fi\let\next=\r@fs\fi\next}
%

%
\newwrite\ffile\global\newcount\figno \global\figno=1
\def\fig{fig.~\the\figno\nfig}
\def\nfig#1{\xdef#1{fig.~\the\figno}%
\writedef{#1\leftbracket fig.\noexpand~\the\figno}%
\ifnum\figno=1\immediate\openout\ffile=figs.tmp\fi\chardef\wfile=\ffile%
\immediate\write\ffile{\noexpand\medskip\noexpand\item{Fig.\ \the\figno. }
\reflabeL{#1\hskip.55in}\pctsign}\global\advance\figno by1\findarg}
\def\xfig{\expandafter\xf@g}\def\xf@g fig.\penalty\@M\ {}
\def\figs#1{figs.~\f@gs #1{\hbox{}}}
\def\f@gs#1{\edef\next{#1}\ifx\next\em@rk\def\next{}\else
\ifx\next#1\xfig #1\else#1\fi\let\next=\f@gs\fi\next}
\newwrite\lfile
{\escapechar-1\xdef\pctsign{\string\%}\xdef\leftbracket{\string\{}
\xdef\rightbracket{\string\}}\xdef\numbersign{\string\#}}

\def\writestop{\def\writestoppt{\immediate\write\lfile{\string\pageno%
\the\pageno\string\startrefs\leftbracket\the\refno\rightbracket%
\string\def\string\secsym\leftbracket\secsym\rightbracket%
\string\secno\the\secno\string\meqno\the\meqno}\immediate\closeout\lfile}}
\def\writestoppt{}\def\writedef#1{}
\def\seclab#1{\xdef #1{\the\secno}\writedef{#1\leftbracket#1}\wrlabeL{#1=#1}}
\def\subseclab#1{\xdef #1{\secsym\the\subsecno}
\writedef{#1\leftbracket#1}\wrlabeL{#1=#1}}
\def\subsubseclab#1{\xdef #1{\secsym\the\subsecno.\the\subsubsecno}
\writedef{#1\leftbracket#1}\wrlabeL{#1=#1}}

\newwrite\tfile \def\writetoca#1{}
\def\leaderfill{\leaders\hbox to 1em{\hss.\hss}\hfill}
\def\writetoc{\immediate\openout\tfile=toc.tmp
   \def\writetoca##1{{\edef\next{\write\tfile{\noindent ##1
   \string\leaderfill {\noexpand\number\pageno} \par}}\next}}}

%
\catcode`\@=12 
%
\edef\tfontsize{\ifx\answ\bigans scaled\magstep3\else scaled\magstep4\fi}
\font\titlerm=cmr10 \tfontsize \font\titlerms=cmr7 \tfontsize
\font\titlermss=cmr5 \tfontsize \font\titlei=cmmi10 \tfontsize
\font\titleis=cmmi7 \tfontsize \font\titleiss=cmmi5 \tfontsize
\font\titlesy=cmsy10 \tfontsize \font\titlesys=cmsy7 \tfontsize
\font\titlesyss=cmsy5 \tfontsize \font\titleit=cmti10 \tfontsize
\skewchar\titlei='177 \skewchar\titleis='177 \skewchar\titleiss='177
\skewchar\titlesy='60 \skewchar\titlesys='60 \skewchar\titlesyss='60
\def\titlefont{\def\rm{\fam0\titlerm}
\textfont0=\titlerm \scriptfont0=\titlerms \scriptscriptfont0=\titlermss
\textfont1=\titlei \scriptfont1=\titleis \scriptscriptfont1=\titleiss
\textfont2=\titlesy \scriptfont2=\titlesys \scriptscriptfont2=\titlesyss
\textfont\itfam=\titleit \def\it{\fam\itfam\titleit}\rm}
 \ifx\answ\bigans\else scaled\magstep1\fi
\ifx\answ\bigans\def\abstractfont{\tenpoint}\else
\font\abssl=cmsl10 scaled \magstep1
\font\absrm=cmr10 scaled\magstep1 \font\absrms=cmr7 scaled\magstep1
\font\absrmss=cmr5 scaled\magstep1 \font\absi=cmmi10 scaled\magstep1
\font\absis=cmmi7 scaled\magstep1 \font\absiss=cmmi5 scaled\magstep1
\font\abssy=cmsy10 scaled\magstep1 \font\abssys=cmsy7 scaled\magstep1
\font\abssyss=cmsy5 scaled\magstep1 \font\absbf=cmbx10 scaled\magstep1
\skewchar\absi='177 \skewchar\absis='177 \skewchar\absiss='177
\skewchar\abssy='60 \skewchar\abssys='60 \skewchar\abssyss='60
\def\abstractfont{\def\rm{\fam0\absrm}
\textfont0=\absrm \scriptfont0=\absrms \scriptscriptfont0=\absrmss
\textfont1=\absi \scriptfont1=\absis \scriptscriptfont1=\absiss
\textfont2=\abssy \scriptfont2=\abssys \scriptscriptfont2=\abssyss
\textfont\itfam=\bigit \def\it{\fam\itfam\bigit}\def\footnotefont{\tenpoint}%
\textfont\slfam=\abssl \def\sl{\fam\slfam\abssl}%
\textfont\bffam=\absbf \def\bf{\fam\bffam\absbf}\rm}\fi
\def\tenpoint{\def\rm{\fam0\tenrm}
\textfont0=\tenrm \scriptfont0=\sevenrm \scriptscriptfont0=\fiverm
\textfont1=\teni  \scriptfont1=\seveni  \scriptscriptfont1=\fivei
\textfont2=\tensy \scriptfont2=\sevensy \scriptscriptfont2=\fivesy
\textfont\itfam=\tenit \def\it{\fam\itfam\tenit}\def\footnotefont{\ninepoint}%
\textfont\bffam=\tenbf \def\bf{\fam\bffam\tenbf}\def\sl{\fam\slfam\tensl}\rm}
\font\ninerm=cmr9 \font\sixrm=cmr6 \font\ninei=cmmi9 \font\sixi=cmmi6
\font\ninesy=cmsy9 \font\sixsy=cmsy6 \font\ninebf=cmbx9
\font\nineit=cmti9 \font\ninesl=cmsl9 \skewchar\ninei='177
\skewchar\sixi='177 \skewchar\ninesy='60 \skewchar\sixsy='60
\def\ninepoint{\def\rm{\fam0\ninerm}
\textfont0=\ninerm \scriptfont0=\sixrm \scriptscriptfont0=\fiverm
\textfont1=\ninei \scriptfont1=\sixi \scriptscriptfont1=\fivei
\textfont2=\ninesy \scriptfont2=\sixsy \scriptscriptfont2=\fivesy
\textfont\itfam=\ninei \def\it{\fam\itfam\nineit}\def\sl{\fam\slfam\ninesl}%
\textfont\bffam=\ninebf \def\bf{\fam\bffam\ninebf}\rm}
%
%

\hyphenation{anom-aly anom-alies coun-ter-term coun-ter-terms}
\def\inv{^{\raise.15ex\hbox{${\scriptscriptstyle -}$}\kern-.05em 1}}

\def\Dsl{\,\raise.15ex\hbox{/}\mkern-13.5mu D} 
\def\dsl{\raise.15ex\hbox{/}\kern-.57em\partial}
\def\del{\partial}

\font\bigit=cmti10 scaled \magstep1
\def\lspace{\ifx\answ\bigans{}\else\qquad\fi}
\def\lbspace{\ifx\answ\bigans{}\else\hskip-.2in\fi} 
\def\boxeqn#1{\vcenter{\vbox{\hrule\hbox{\vrule\kern3pt\vbox{\kern3pt
        \hbox{${\displaystyle #1}$}\kern3pt}\kern3pt\vrule}\hrule}}}
\def\mbox#1#2{\vcenter{\hrule \hbox{\vrule height#2in
                \kern#1in \vrule} \hrule}}  
%

\def\darr#1{\raise1.5ex\hbox{$\leftrightarrow$}\mkern-16.5mu #1}

\overfullrule=0pt
\abovedisplayskip=12pt plus 3pt minus 3pt
\belowdisplayskip=12pt plus 3pt minus 3pt
\sequentialequations
%
%
%
\message{S-Tables Macro v1.0, ACS, TAMU (RANHELP@VENUS.TAMU.EDU)}
%
%
\newhelp\stablestylehelp{You must choose a style between 0 and 3.}%
\newhelp\stablelinehelp{You
should not use special hrules when stretching
a table.}%
\newhelp\stablesmultiplehelp{You have tried to place an S-Table
inside another
S-Table.  I would recommend not going on.}%
%
%
\newdimen\stablesthinline
\stablesthinline=0.4pt
\newdimen\stablesthickline
\stablesthickline=1pt
%
%
\newif\ifstablesborderthin
\stablesborderthinfalse
\newif\ifstablesinternalthin
\stablesinternalthintrue
\newif\ifstablesomit
\newif\ifstablemode
\newif\ifstablesright
\stablesrightfalse
%
%
\newdimen\stablesbaselineskip
\newdimen\stableslineskip
\newdimen\stableslineskiplimit
%
%
\newcount\stablesmode
\newcount\stableslines
\newcount\stablestemp
\stablestemp=3
\newcount\stablescount
\stablescount=0
\newcount\stableslinet
\stableslinet=0
%
%
%
\newcount\stablestyle
\stablestyle=0
%
%
\def\stablesleft{\quad\hfil}%
\def\stablesright{\hfil\quad}%
%
%
\catcode`\|=\active%
%
%
\newcount\stablestrutsize
\newbox\stablestrutbox
\setbox\stablestrutbox=\hbox{\vrule height10pt depth5pt width0pt}
\def\stablestrut{\relax\ifmmode%
                         \copy\stablestrutbox%
                       \else%
                         \unhcopy\stablestrutbox%
                       \fi}%
%
%
\newdimen\stablesborderwidth
\newdimen\stablesinternalwidth
\newdimen\stablesdummy
\newcount\stablesdummyc
\newif\ifstablesin
\stablesinfalse
%
%
\def\begintable{\stablestart%
  \stablemodetrue%
  \stablesadj%
  \halign%
  \stablesdef}%
\def\stablesadj{%
  \ifcase\stablestyle%
    \hbox to \hsize\bgroup\hss\vbox\bgroup%
  \or%
    \hbox to \hsize\bgroup\vbox\bgroup%
  \or%
    \hbox to \hsize\bgroup\hss\vbox\bgroup%
  \or%
    \hbox\bgroup\vbox\bgroup%
  \else%
    \errhelp=\stablestylehelp%
    \errmessage{Invalid style selected, using default}%
    \hbox to \hsize\bgroup\hss\vbox\bgroup%
  \fi}%
\def\stablesend{\egroup%
  \ifcase\stablestyle%
    \hss\egroup%
  \or%
    \hss\egroup%
  \or%
    \egroup%
  \or%
    \egroup%
  \else%
    \hss\egroup%
  \fi}%
\def\stablestart{%
  \ifstablesin%
    \errhelp=\stablesmultiplehelp%
    \errmessage{An S-Table cannot be placed within an S-Table!}%
  \fi
  \global\stablesintrue%
  \global\advance\stablescount by 1%
  \message{S-Tables Generating Table \number\stablescount}%
  \begingroup%
  \stablestrutsize=\ht\stablestrutbox%
  \advance\stablestrutsize by \dp\stablestrutbox%
  \ifstablesborderthin%
    \stablesborderwidth=\stablesthinline%
  \else%
    \stablesborderwidth=\stablesthickline%
  \fi%
  \ifstablesinternalthin%
    \stablesinternalwidth=\stablesthinline%
  \else%
    \stablesinternalwidth=\stablesthickline%
  \fi%
  \tabskip=0pt%
  \stablesbaselineskip=\baselineskip%
  \stableslineskip=\lineskip%
  \stableslineskiplimit=\lineskiplimit%
  \offinterlineskip%
  \def\borderrule{\vrule width \stablesborderwidth}%
  \def\internalrule{\vrule width \stablesinternalwidth}%
  \def\thinline{\noalign{\hrule height \stablesthinline}}%
  \def\thickline{\noalign{\hrule height \stablesthickline}}%
  \def\trule{\omit\leaders\hrule height \stablesthinline\hfill}%
  \def\ttrule{\omit\leaders\hrule height \stablesthickline\hfill}%
  \def\tttrule##1{\omit\leaders\hrule height ##1\hfill}%
  \def\stablesel{&\omit\global\stablesmode=0%
    \global\advance\stableslines by 1\borderrule\hfil\cr}%
  \def\el{\stablesel&}%
  \def\elt{\stablesel\thinline&}%
  \def\eltt{\stablesel\thickline&}%
  \def\elttt##1{\stablesel\noalign{\hrule height ##1}&}%
  \def\elspec{&\omit\hfil\borderrule\cr\omit\borderrule&%
              \ifstablemode%
              \else%
                \errhelp=\stablelinehelp%
                \errmessage{Special ruling will not display properly}%
              \fi}%
  \def\stmultispan##1{\mscount=##1 \loop\ifnum\mscount>3
\stspan\repeat}%
  \def\stspan{\span\omit \advance\mscount by -1}%
  \def\multicolumn##1{\omit\multiply\stablestemp by ##1%
     \stmultispan{\stablestemp}%
     \advance\stablesmode by ##1%
     \advance\stablesmode by -1%
     \stablestemp=3}%
  \def\multirow##1{\stablesdummyc=##1\parindent=0pt\setbox0\hbox\bgroup%
    \aftergroup\emultirow\let\temp=}
  \def\emultirow{\setbox1\vbox to\stablesdummyc\stablestrutsize%
    {\hsize\wd0\vfil\box0\vfil}%
    \ht1=\ht\stablestrutbox%
    \dp1=\dp\stablestrutbox%
    \box1}%

\def\stpar##1{\vtop\bgroup\hsize ##1%
     \baselineskip=\stablesbaselineskip%
     \lineskip=\stableslineskip%

\lineskiplimit=\stableslineskiplimit\bgroup\aftergroup\estpar\let\temp=}%
  \def\estpar{\vskip 6pt\egroup}%
  \def\stparrow##1##2{\stablesdummy=##2%
     \setbox0=\vtop to ##1\stablestrutsize\bgroup%
     \hsize\stablesdummy%
     \baselineskip=\stablesbaselineskip%
     \lineskip=\stableslineskip%
     \lineskiplimit=\stableslineskiplimit%
     \bgroup\vfil\aftergroup\estparrow%
     \let\temp=}%
  \def\estparrow{\vfil\egroup%
     \ht0=\ht\stablestrutbox%
     \dp0=\dp\stablestrutbox%
     \wd0=\stablesdummy%
     \box0}%
  \def|{\global\advance\stablesmode by 1&&&}%
  \def\|{\global\advance\stablesmode by 1&\omit\vrule width 0pt%
         \hfil&&}%
  \def\vt{\global\advance\stablesmode by 1&\omit\vrule width
\stablesthinline%
          \hfil&&}%
  \def\vtt{\global\advance\stablesmode by 1&\omit\vrule width
\stablesthickline%
          \hfil&&}%
  \def\vttt##1{\global\advance\stablesmode by 1&\omit\vrule width ##1%
          \hfil&&}%
  \def\vtr{\global\advance\stablesmode by 1&\omit\hfil\vrule width%
           \stablesthinline&&}%
  \def\vttr{\global\advance\stablesmode by 1&\omit\hfil\vrule width%
            \stablesthickline&&}%
  \def\vtttr##1{\global\advance\stablesmode by 1&\omit\hfil\vrule
width ##1&&}%
  \stableslines=0%
  \stablesomitfalse}
\def\stablesdef{\bgroup\stablestrut\borderrule##\tabskip=0pt plus 1fil%
  &\stablesleft##\stablesright%
  &##\ifstablesright\hfill\fi\internalrule\ifstablesright\else\hfill\fi%
  \tabskip 0pt&&##\hfil\tabskip=0pt plus 1fil%
  &\stablesleft##\stablesright%
  &##\ifstablesright\hfill\fi\internalrule\ifstablesright\else\hfill\fi%
  \tabskip=0pt\cr%
  \ifstablesborderthin%
    \thinline%
  \else%
    \thickline%
  \fi&%
}%
\def\endtable{\advance\stableslines by 1\advance\stablesmode by 1%
   \message{- Rows: \number\stableslines, Columns:
\number\stablesmode>}%
   \stablesel%
   \ifstablesborderthin%
     \thinline%
   \else%
     \thickline%
   \fi%
   \egroup\stablesend%
\endgroup%
\global\stablesinfalse}
%

\def\bar{\overline}

\def\ads{{$AdS_5$}}

\def\adst{{$AdS_5\times T^{nn}$}}

\def\Tpq{$T^{nn'}$}

\def\half{{1\over2}}
\def\bz{{\bar z}}
\def\Osn{{\cal O}_{s,p}}
\def\bOsn{\overline{{\cal O}}_{s,p'}}
\def\Osmn{{\cal O}_{s-1,p}}
\def\bOsmn{\overline{{\cal O}}_{s-1,p'}}
\def\kk{Kaluza-Klein}
%


\lref\adswit{E. Witten, {\it Anti-de Sitter Space and Holography},
hep-th/9802150.}
\lref\mikseif{For a review see Michael R. Douglas and S.  Randjbar-Daemi,
{\it Two Lectures on AdS/CFT Correspondence}, hep-th/9902022.}
\lref\klebwit{I. R. Klebanov and E. Witten, {\it Superconformal Field Theory
on Threebranes at a Calabi-Yau Singularity}, hep-th/9807080.}
\lref\pope{D. N Page and  C.N. Pope,{\it Which Compactifications
of D=11 Supergravity are Stable}, Phys.Lett. B144 (1984) 346;
P. Candelas and X. de la Ossa,{\it Comments on Conifolds},
Nucl. Phys. {\bf B342} (1990) 246.}
\lref\witten{E. Witten, {\it Ground Ring of Two Dimensional String Theory},
hep-th/9108004;
E. Witten and B. Zwiebach, {\it Algebraic Structures and Differential
Geometry in 2D String Theory}, hep-th/9201056.}
\lref\gjm{D. Ghoshal, D. P. Jatkar and S. Mukhi, {\it Kleinian Singularities
and the Ground Ring of $c=1$ String Theory}, hep-th/9206080.}
\lref\gkp{S. S. Gubser, I. R. Klebanov and A. M. Polyakov, {\it Gauge Theory
Correlators from Non-critical String Theory}, hep-th/9802109.}
\lref\malda{J. Maldacena, {\it The Large $N$ Limit of Superconformal Field
Theories and Supergravity}, hep-th/9711200.}
\lref\gub{S. S. Gubser, {\it Einstein Manifolds and Conformal Field Theories},
hep-th/9807164.}
\lref\gv{D. Ghoshal and C. Vafa, {\it c=1 string as the Topological Theory
of the Conifold}, Nucl. Phys. {\bf B453}(1995) 121.}
\lref\ag{I. Antoniadis, E. Gava, K.S. Narain and T. R. Taylor,
{\it  N=2 type II-heterotic duality and the Higher derivative F-terms},
hep-th/9507115.}
\lref\gvafa{ Rajesh Gopakumar and C. Vafa, {\it  Topological gravity as
Large N Topological gauge Theory}, hep-th/9802016 ,
{\it M-Theory and Topological strings I and II}, hep-th/ 9809187 and
hep-th/ 9812127 and {\it On the Gauge Theory/Geometry
Correspondence}, hep-th/9811131.}
\lref\rss{S. Randjbar-Daemi, A. Salam and J.  Strathdee, {\it Spontaneous
Compactification
in Six Dimensional Einstein - Maxwell Theory}, Nucl. Phys. {\bf B214}(1983)
491.}
\lref\krn{H. J. Kim, L. J. Romans and P. van Nieuwenhuizen, {\it Mass spectrum
of chiral ten-dimensional N=2 supergravity on $S^5$}, Phys. Rev. {\bf D32}
(1985) 389.}

{\nopagenumbers
\Title{\vtop{\hbox{hep-th/9904187}
\hbox{IC/99/42}
\hbox{MRI-PHY/990409}
}}
{\vbox{\hbox{\centerline{Type IIB string theory on \ads $\times$\Tpq }}
}}
\centerline{Dileep P. Jatkar$^{*\dagger}$ and S. Randjbar-Daemi$^*$
\foot{E-mail: dileep@mri.ernet.in, seif@ictp.trieste.it}}
\vskip 6pt
\centerline{$^*$\it Abdus Salam International Centre for Theoretical Physics,}
\centerline{\it Strada Costiera 11, Miramare, Trieste 34100, ITALY}
\vskip 3pt
\centerline{$^\dagger$\it Mehta Research Institute of Mathematics and
Mathematical Physics,\foot{Permanent Address}}
\centerline{\it Chhatnag Road, Jhusi, Allahabad 211 019, INDIA}
\medskip
\centerline{ABSTRACT}

We study \kk\ spectrum of type IIB string theory compactified on
$AdS_5 \times T^{nn'}$  in the context of $AdS/CFT$ correspondence.
We examine some of the modes of the complexified $2$ form potential as an
example and show that for the states at the bottom of the  \kk\ tower
 the corresponding $d=4$ boundary field operators have  rational conformal
dimensions.  The masses of some of the fermionic modes in the bottom of
each tower as functions of the $R$ charge in the boundary conformal theory
are also rational. Furthermore the modes in the bottom of the towers
originating from $q$ forms on $T^{11}$  can be put in correspondence with the
BRS cohomology classes of the $c=1$ non critical string theory with ghost
number $q$. However, a more detailed investigation is called for, to
clarify further the relation of this supergravity background with the
$c=1$ strings.
\vfill
\leftline{March 1999}
\eject}
\ftno=0
\newsec{Introduction}

It has recently been suggested by Klebanov and Witten\refs\klebwit\ that
the world volume super Yang Mills theory  of parallel $D_3$-branes near a
conical singularity of a Calabi-Yau manifold are related to string theory on
$AdS_5\times T^{11}$. They argue that string theory on
\ads $\times T^{11}$,  with $T^{11} = (SU(2)\times SU(2))/U(1)$
can be described in terms of an
${\cal N}=1$ superconformal $SU(N)\times SU(N)$ gauge theory.  In the absence
of a rigorous proof of the  AdS/CFT correspondence \refs\malda\gkp\
one needs to have the
\kk\ spectrum of the supergravity in the above geometry. The \kk\ masses are
used to identify the anomalous dimensions of the operators on the Yang
Mills side\refs\adswit\mikseif .

The manifold $T^{11}$ also enters in the geometrical description of
$c=1$ matter compactified on a circle at the self dual radius coupled to the
$d=2$ gravity, the so called non critical d=2  string theory. It has been
shown in \refs\witten\ that the symmetry algebra of the
volume preserving diffeomorphisms of the cone based on $T^{11}$ is in fact
in a $1-1$ correspondence with the symmetry algebra of (1,0) and (0,1)
conformal primary fields of the $c=1$ theory at the self dual
radius. It is therefore tempting to ask if there is any
relationship between the $c=1$ theory and the type IIB string theory
compactified on \ads $\times T^{11}$. We will also touch upon the relation
of \kk\ spectrum on $T^{nn}$ and $c=1$ string theory at $n$-times the self
dual radius in the last section.

The \kk\ modes are characterised by a set of quantum numbers  $\{l, l', s\}$,
where $l$ and $l'$ characterise the $SU(2)\times SU(2)$ multiplets and $s$,
related to the $R$ charge in the Yang Mills side, is a $U(1)$ charge taking
integer or $1/2$ integer values. The lowest values of $l$ and $l'$ will depend
on $s$.  For any given $s$ there is an infinite tower of Kaluza Klein modes.
We shall  show that for any given $s$ and for some ( but not all) of the modes
with the smallest possible values of $l$ and $l'$ there  corresponds well
defined cohomology classes in the ghost numbers 0, 1 and
2 sectors of the $c=1$ theory.  More precisely we shall show that
the Kaluza Klein modes in $AdS_5$  originating from $q$
forms in $T^{11}$ are in correspondence
with the cohomology classes with ghost number $q$ on the $c=1$ side. They have
identical $SU(2)\times SU(2)$ quantum numbers. However the modes
originating from the components of the metric in the internal space seem to
be unmatched. Along the way  we shall show that the derivation of the \kk\
spectrum can be reduced to the problem of the diagonalization of differential
operators acting on tensors and spinors in the background of monopole
fields on $S^2\times S^2$. As an example we give the \kk\ masses of some of
the bosonic and fermionic fields.

Possible correspondence between $c=1$ theory at self dual radius and the type
IIB strings on conifold singularities have been noted in the past by Ghoshal
and Vafa \refs\gv\ . We shall comment on this paper at the end the present
note.

\newsec{The manifolds $T^{nn'}$  and the Kaluza-Klein modes}

Although manifolds $T^{nn}$ are related to the $c=1$ string theory, \kk\
spectrum can be studied on any $T^{nn'}$ for $n\not= n'$. Below we will
study the \kk\ spectrum of Type IIB string theory compactified on $T^{nn'}$.
In the end when we will relate our results to $c=1$ string theory we will
take $n=n'$.

The  geometry of the manifolds $T^{nn'}$ is defined in terms of the
metric\refs\pope
\eqn\one{ds^4=c^2(dy_5-n\cos y_1dy_2-n'\cos y_3dy_4)^2+
a^2(dy_1^2+\sin^2y_1dy_2^2)+a^{{\prime}^2}(dy_3^2+\sin^2y_3dy_4^2).}
All the coordinates $y^{\alpha}$ are angles, such that $ y=({y^1}, y^{2})$
parametrise a $S^2$ of radius $a$ while $y' = (y^3 , y^4)$ parametrise a
$S^2$ of radius $a'$. The angle $y^5$ ranges from 0 to $4\pi$. The constant
$c$ is the radius of the circle defined by $y^5$. The constants $n$ and $n'$
will be taken to be integers. We shall denote these manifolds by
$T^{nn^{\prime}}$. Locally they look like $S^2\times S^2\times S^1$. Their
isometry group is $SU(2)\times SU(2)\times U(1)$, where the $U(1)$ factor
is due to the translational invariance of the coordinate $y^5$. It has been
argued by Klebanov and Witten that the $U(1)$ factor should be identified with
 the $R$ symmetry of the world volume ${\cal N}=1$, $d=4$ superconformal
field theory which arises as a consequence of the $AdS/CFT$ correspondence.
We shall see that it can
also be put in correspondence with the $U(1)$ group generated by the Liouville
mode of the $c=1$ string theory. In this way the $R$ charges of the boundary
$d=4$ superconformal theory will be set in correspondence with the Liouville
momenta of the $c=1$ theory.

In order for $AdS_5\times T^{nn'}$ to be a supersymmetric solution of the type
IIB field equations it is necessary that
$a=a'= {1\over {\sqrt 6 \mid e\mid}}$, $n=\pm n'$ and $ec=-{1\over {3n}}$,
where, $e^2$ is related to the $AdS_5$ cosmological constant
through $R_{\mu\nu}= -4e^2g_{\mu\nu}$.  For the present discussion
we can keep the background parameters arbitrary.

Our strategy for the Kaluza Klein expansion is to dispose of the $y^5$
coordinate by Fourier expansion. As a result of this we obtain infinite
number of fields living on $AdS_5\times S^2\times S^2$.  Writing every
object in an orthonormal basis of the $ S^2\times S^2$ manifold we obtain
infinite number of fields coupled to a magnetic monopole field of charge
$ns$ on the first $S^2$ and $n^{\prime}s$  on the second $S^2$. The harmonic
expansion on the magnetic monopole background has been studied in detail in
\refs\rss\ . We will use the formalism of this reference to write down the
eigenvalues of the $ S^2\times S^2$ Laplacian acting on any field. In this
way we will obtain the Kaluza Klein modes which depend on the coordinates
of $AdS_5$ only.

To implement this idea we need the components of the $5$-bein
and the spin connections on $T^{nn'}$. They are given by
\eqn\viel{\eqalign{e_{\underline1}^1&={1\over a },\quad e_{\underline  2}^2=
{1\over {a \sin y_1} },\quad e_{\underline 2}^5= {n\over {a \cot y_1}},\cr
e_{\underline3}^3 &= {1\over a },\quad e_{\underline  4}^4= {1\over {a'
\sin y_3} },\quad
e_{\underline 4}^5= {n^{\prime}\over {a^{\prime} \cot y_3}},\quad
e_{\underline5}^5= {1\over c }.}}
{}From these one can calculate the components of the spin connections. They
are given by
\eqn\spinc{\eqalign{\omega_{\underline a[\underline b\underline c]}&=
\omega_{\underline a}\varepsilon_{\underline b\underline c},\quad
\omega_{\underline 5[\underline a\underline b]}=
e\varepsilon_{\underline a\underline b},\quad \omega_{\underline a[\underline
b
\underline 5]}=
-e\varepsilon_{\underline a\underline b}\cr
\omega_{\underline m[\underline n\underline l]}&= \omega_{\underline m}
\varepsilon_{\underline n\underline l},\quad  \omega_{\underline 5[\underline
m
\underline n]}=
e\varepsilon_{\underline m\underline n},\quad \omega_{\underline m[\underline
n
\underline 5]}=
-e\varepsilon_{\underline m\underline n},}}
where  $\omega_{\underline a}= (\omega_{\underline 1}=0,
\omega_{\underline 2}= -{1\over a}\cot y_1)$ and
$\omega_{\underline m}= (\omega_{\underline 3}=0, \omega_{\underline 3}=
-{1\over a'}\cot y_3)$. These are
just the components of a magnetic monopole potential on each $S^2$. For
what follows
it is important to note that  $e_{\underline a}^5= -n \omega_{\underline a}$
and  $e_{\underline m}^5= -n' \omega_{\underline m}$.

To clarify the scheme let us look at the derivatives of a $d=10$ scalar
$\Phi(x, y^5, y, y')$.
First we write
\eqn\four{\Phi(x, y^5, y, y') = \sum_{s \in \half Z} e ^{isy^5}\Phi^{s}(x,
y, y')}
Since the angle $y^5$ ranges over a period of $4\pi$, $s$ takes integer as
well as 1/2 integer
values. The derivative operator on \Tpq\ then acts on the Fourier
modes $\Phi^s$
according to
\eqn\five{\eqalign{D_{\underline 5}\Phi(x, , y, y') &= {is\over c}
\Phi(x, y, y')\cr
D_{\underline a}\Phi^s(x, y, y') &= (\del_{\underline a} - i ns
\omega_{\underline a})\Phi^s(x, y,
y') = \nabla_{\underline a} \Phi^s(x, y, y') \cr
D_{\underline m}\Phi^s(x, y, y') &=(\del_{\underline m} - i n^{\prime }s
\omega_{\underline m})\Phi^s(x,
y, y') =\nabla_{\underline m} \Phi^s(x, y, y').}}
We thus see that $\nabla_{\underline a} \Phi^s(x, y, y') $ and
$\nabla_{\underline m}\Phi^s(x, y, y')$ are precisely the
orthonormal frame components of the covariant derivatives of a charged
scalar field coupled to  $U(1)$ monopole fields on each
sphere. Furthermore the $U(1)$ charge of $\Phi^s$ on the first $S^2$ is
$ns$ and on the second $S^2$ it is $n' s$.

The above covariant derivatives dictate the form of
harmonic expansions for each $\Phi^s$.
Following the notation of \refs\rss\ we can write
\eqn\six{\Phi^s(x, y, y')= \sum_{l\ge \mid n_{1}s\mid, \mid \lambda \mid\le l}
(2l+1)^{1\over {2}}
\sum_{l'\ge \mid n's\mid, \mid\lambda\mid \le l} (2l' +1)^{1\over {2}}
 \Phi^{s,l, l'}_{\lambda,\lambda'}(x) D^l_{ns, \lambda}(L_y^{-1})
D^{l'}_{ns, \lambda'} (L_{y'}^{-1})}
In these formulae the spheres are regarded as the coset
manifolds $SU(2)/U(1)$. The  $SU(2)$ group element $L_y$
represents the point $y \in S^2$ and $ D^l_{ns, \lambda}(L_y^{-1})$ are
the matrix elements of
 the unitary irreducible representations
of $SU(2)$ carrying spin $l$. Their most important property for us is that
\eqn\seven{\nabla_{\pm}  D^l_{q, \lambda}(L_y^{-1})={i\over{\sqrt{2}a_1}}\sqrt
{(l\mp q)( l\pm q +1)}D^l_{q, \lambda}(L_y^{-1})}
 The $U(1)$ basis $\pm$  on the first $S^2$ are defined by
$
\nabla_{\pm} ={1\over {\sqrt 2}} (\nabla_{\underline 1} \mp
i\nabla_{\underline 2}).
$
On the second $S^2$ we shall denote  the $U(1)$ basis by a prime, viz,
$
\nabla^{\prime}_{\pm} ={1\over{\sqrt 2}} (\nabla_{\underline 3} \mp
i\nabla_{\underline 4}).
$

{}From these relations we can easily read the eigenvalues of the Laplacian
action on the charge q objects on $S^2$, viz,
\eqn\ten{\nabla_{\underline a}\nabla_{\underline a}D^l_{q, \lambda}(L_y^{-1})=
(\nabla_{+}\nabla_{-} +\nabla_{-}\nabla_{+} ) D^l_{q, \lambda}(L_y^{-1})=
-{1\over a^2} ( l(l+1)-q^2)  D^l_{q, \lambda}(L_y^{-1})}
The eigenvalues of the $5$ dimensional Laplacian on $\Phi^s$ can now be
read immediately.  As an example let us
consider the complex dilaton. The linearised equation for this field is
given by
\eqn\eleven{(D_\mu D^\mu +D_\alpha D^\alpha)\Phi(x, y^5, y, y')=0}
Noting that
\eqn\scalone{D_\alpha D^\alpha = -({s\over c})^2 + \nabla_{\underline a}
\nabla_{\underline a} + \nabla_{\underline m}\nabla_{\underline m}}
we can simply read the $AdS_5$ mass of the mode
$\Phi^{s,l, l'}_{\lambda,\lambda'}(x)$ as
\eqn\scaltwo{
m^2=  {s^2\over c^2}+{1\over a^2}(l(l+1)-(ns)^2) + {1\over a^{\prime 2}}
(l'(l'+1)-(n' s)^2) }
where $l\ge \mid ns\mid$ and $l'\ge \mid n^{\prime}s\mid$.

For the next example we consider the  complex, 2-form potential $B_{MN}$.
Firstly, The components $B_{\mu\nu}$ are \Tpq\ scalars. Therefore they must
be treated exactly in the same way as $\Phi$. The remaining components can
be decomposed  into various tensors on $S^2\times S^2$ . Here we list the
result,
\eqn\lapone{D_{\alpha}D_{\alpha} B_{\mu\underline 5}= [-({s\over c})^2 +
\nabla_{\underline a}\nabla_{\underline a} +
 \nabla_{\underline m}\nabla_{\underline m} - 4e^2]B_{\mu\underline 5}-
2e\varepsilon_{\underline a\underline b}\nabla_{\underline a}
B_{\mu\underline b}-2e\varepsilon_{\underline m\underline n}
\nabla_{\underline m} B_{\mu\underline n} }
\eqn\laptwo{D_{\alpha}D_{\alpha} B_{\mu\pm}= [-({s\over c} \mp e)^2 +
\nabla_{\underline a}\nabla_{\underline a} +
 \nabla_{\underline m}\nabla_{\underline m} - e^2]B_{\mu\pm}  \mp  2ie
\nabla_{\pm}B_{\mu\underline 5}}
\eqn\lapthree{D_{\alpha}D_{\alpha} B'_{\mu\pm}= [-({s\over c} \mp
e)^2 + \nabla_{\underline a}\nabla_{\underline a} +
 \nabla_{\underline m}\nabla_{\underline m} - e^2]B'_{\mu\pm}  \mp  2ie
\nabla'_{\pm} B_{\mu\underline 5}}
In these relations the covariant derivatives are defined by the $U(1)$
charge of each object. For  an object $B_{pq}$ of charge $p$ on the first
$S^2$ and charge $q$ in the second $S^2$ we have,
$$
\nabla_{\underline a}B_{pq} =(\del _{\underline a} -i p
\omega_{\underline a})B_{pq}\quad {\rm and} \quad
\nabla_{\underline m}B_{pq} =(\del _{\underline m} -i q\omega_{\underline m})
B_{pq}
$$
where $\del_{\underline a}$ and $\del_{\underline m}$  denote the partial
derivatives in the orthonormal basis. The
 $(p,q)$ charges of $B_{\mu \underline 5}$, $B_{\mu \pm}$, $B'_{\mu
\pm}$, and
$B_{\pm (\pm)^{\prime}}$  are, respectively , $(ns, n's)$, $(ns\pm 1, n's)$,
$(ns, n's\pm 1)$ and $( ns\pm 1, n' s \pm 1)$. These charges determine the
lower bounds of $l$ and $l'$
in the harmonic expansion of $B_{pq}$ on $S^2\times S^2$. The
expansion of $B_{pq}$
is identical to the one of $\Phi$ given above except that we should put the
lower bounds $l\ge\mid p\mid$ and $l'\ge\mid q\mid$.

Other bosonic fields can be treated in a similar way. We tabulate  the $(p,q)$
charges of all of the bosonic fields of the type IIB supergravity in
table I.

The technique outlined above can be used to obtain the Kaluza Klein masses
of small perturbations around the background solution \adst .
The full analysis, although straightforward, is quite lengthy and will not
be given here. Here we shall give the result of calculation
of the masses of those modes which decouple without too much labour from
the rest of the spectrum. The easiest one is the complex scalar
for which we gave the spectrum of the masses in the
previous section. We shall next consider the complex 2-form potential
$B_{MN}$.

Note that for each $s$ there is an infinite tower of modes in each field. We
shall consider only some of these towers whose mass can easily be deduced.
For each $s$ this happens for the modes with the smallest values of $l$ and
$l'$. Consider first the modes with $ns\ge 1$.  In this case the modes
$B^{s, ns-1,ns}_{\mu -}$ decouple from the rest of the system. Using the
background values of
$
 {1\over a^2}= 6e^2, {1\over {ec}}= -3n
$
we obtain
\eqn\twentyone{2D^\mu D_{[\mu}B_{\nu]-}^{s, ns-1,ns}- e^2[(3ns +1)^2 -1]
B_{\nu-}^{s,ns-1,ns} =0}
Identical equation will result for $B_{\nu-}^{\prime s, ns,ns-1}$.

For the range of $ns\le -1$ it is the leading modes of $B_{\nu+}^{s,-
ns-1, -ns}$
and $B_{\nu+}^{\prime s,-ns,-ns-1}$ which decouple. Their equation of motion
becomes
\eqn\twentytwo{2D^\mu D_{[\mu}B_{\nu]+}^{s,-ns-1,- ns}- e^2[(3ns -1)^2 -1]
B_{\nu+}^{s,-ns-1,- ns} =0}
and an identical equation for $B_{\nu+}^{\prime s,-ns,-ns-1}$.

Following  \refs\krn\ we shall identify  $e^2[(3ns +1)^2 -1]$ as the $AdS$
$mass^2$
of $B^{s, ns-1,ns}_{\mu -}$ and  $B_{\nu-}^{\prime s, ns,ns-1}$. Likewise
$e^2[(3ns -1)^2 -1]$ will be identified with the $AdS$ mass of
$B_{\nu+}^{s,- ns-1, -ns}$
and $B_{\nu+}^{\prime s,-ns,-ns-1}$.

According to the $AdS/CFT$ correspondence every bulk field in $AdS_5$
should correspond to a well defined gauge invariant operator
in the boundary $d=4$ theory.  For an $AdS$ mode of mass $m$ which
originates from a $p$ form
field in the internal space  the conformal dimension $\Delta$ of the $d=4$
theory is given by $(\Delta-p)(\Delta+p-4)=m^2$. Clearly only very small
subset of modes will produce a rational solution for $\Delta$.  This
is what happens to the modes singled out above, namely, the conformal
dimension of the fields dual to $B^{s, ns-1,ns}_{\mu -}$ and
$B_{\nu-}^{\prime s, ns,ns-1}$
turn out to be $3(1+ns)$ and those of the fields corresponding to
$B^{s, ns-1,ns}_{\mu +}$ and  $B_{\nu +}^{\prime s, ns,ns-1}$ are equal to
$3ns+1$.

The same phenomenon also happens for the complex dilaton modes.
This was observed by Gubser \refs\gub\ . From our equation \scaltwo\ it
follows that
for any given $s$ the smallest masses are obtained for $l= l'=\mid ns\mid$,
which is  $m^2= e^2[(3\mid ns\mid +2)^2 -4].$
Plugging this in the formula for the dimension with $p=0$ produces the
positive root
$
\Delta= 4+3\mid ns\mid .
$
Gubser has conjectured that for a given $s$ this will happen to every mode
in the bottom of the tower with that value of $s$.

We can also obtain the fermion masses. The easiest one is the dilatino. It
satisfies the equation
\eqn\dilatino{ (D\!\!\!\!/_x  + iD\!\!\!\!/_y  + e)\chi =0}

We categorise the spectrum of $iD\!\!\!\!/_y$ using the quantum number $s$.
For the sake of simplicity let us set $n=n'=1$.
Let us take $\mid s\mid\ge 1/2$. In this case we have four subsectors.
We will use the notation $\epsilon(s)$ to denote sign of $s$ and define
$\epsilon (0)=1$ in the fourth sector below.  We shall also introduce
the function $\nu(l,s)$ through
$\nu(l,s)= {1\over a}\sqrt{(l+{1\over 2})^2- s^2}$
and define $\nu'(l',s) =\nu (l', s)$.

In the first sector we have only one eigenvalue
\eqn\secone{\lambda(l,l')=\lambda(-\half+\mid s\mid, -\half +\mid s\mid)=
-\epsilon(s)e(1+3\mid s\mid).}
The second sector  is defined by $l=\mid s\mid -1/2$, $l'\ge \mid s\mid +1/2$.
Here we have two towers parametrised by $l'$
\eqn\sectwo{\lambda_\pm(-\half+\mid s\mid, l')=-\epsilon(s){e\over 2}
\pm\sqrt{\nu^{\prime 2}+{e^2\over 4}[1+12\mid s\mid(1+3\mid s\mid)]}}
and in particular for $l'=\mid s\mid +1/2$ the eigenvalue is rational
\eqn\secr{\lambda_\pm(-\half+\mid s\mid, \half +\mid s\mid)=
-\epsilon(s){e\over 2}\pm{\mid e\mid\over 2}(6\mid e\mid+5).}

In the third sector, which is defined by $l\ge \mid s\mid +1/2$ ,
$l'=\mid s\mid -1/2$,  for each $s$ again we have two towers  parametrised
by $l$
\eqn\secthree{\lambda_\pm(l, -\half+\mid s\mid)=-\epsilon(s){e\over 2}
\pm\sqrt{\nu^2+{e^2\over 4}[1+12\mid s\mid(1+3\mid s\mid)]}}
and as in the case of sector two, for $l=\mid s\mid +1/2$ the eigenvalue is
rational. i.e.,
\eqn\ratthree{\lambda_\pm(\half+\mid s\mid, -\half +\mid s\mid)=
-\epsilon(s){e\over 2}\pm{\mid e\mid\over 2}(6\mid e\mid+5).}
In the fourth sector, we will relax the condition to $\mid s\mid\ge 0$. Thus
the $s=0$ modes are included in this sector.  This sector is defined by
 $l\ge \mid s\mid +1/2$ , $l'\ge\mid s\mid +1/2$.  Here we have four towers
and they are given by
\eqn\secfour{\eqalign{\lambda_1^\pm(l, l',s)&=-\epsilon(s){e\over 2}\pm
\sqrt{\nu^2+\nu^{\prime 2}+e^2(3\mid s\mid +\half)^2}\cr
\lambda_2^\pm(l, l',s)&=\epsilon(s){e\over 2}\pm
\sqrt{\nu^2+\nu^{\prime 2}+e^2(3\mid s\mid -\half)^2}.}}
In this case, however, we find that for $l=l'=\mid s\mid+1/2$, only two
eigenvalues, $\lambda_2$ become rational, except for the case $s=0$ when
$\lambda_1$ also becomes rational.
\eqn\ratfour{\eqalign{\lambda_1^\pm(\half +\mid s\mid ,\half +\mid s\mid)&=
-\epsilon(s){e\over 2}\pm{\mid e\mid\over 2}\sqrt{36s^2+108\mid s\mid+49}\cr
\lambda_2^\pm(\half +\mid s\mid ,\half +\mid s\mid)&=\epsilon(s){e\over 2}\pm
{\mid e\mid\over 2}(6\mid s\mid+7).}}

{}From the above list of the eigenvalues of  $iD\!\!\!\!/_y$ one can also
obtain the masses of the $AdS$ gravitinos as well as the $gamma$ trace of
the gravitino, $\gamma^\mu\psi_\mu$.
This latter field satisfies an equation similar to \dilatino\ in
which $e$ is replaced by $3e$, whereas the $D_\mu$ traceless and $gamma$
traceless gravitino $\phi_\mu$ satisfies
$$D\!\!\!\!/_x\phi_{(\mu)}^j-(\lambda^j-e)
\phi_{(\mu)}^j=0.$$
where $\lambda^j$ indicates the eigenvalues of $iD\!\!\!\!/_y$ and $j$ runs
over various sectors.

The fermionic modes in the direction of the Killing spinor in $T^{nn}$ satisfy
equations similar to those in refs\krn\ .

\newsec{Relation to c=1 string theory}

The fundamental fields of the $c=1$ theory are two scalars $X$ and $\phi$
and the $b,c$ ghost fields. $X$ is targeted on a $S^1$ while $\phi$ is
coupled to a background charge. At the self dual radius $R=1/\sqrt 2$ of the
circle there is a $SU(2)\times SU(2)$ symmetry. The BRS cohomology classes
are organised according to the representations of this group. These classes
are also labeled by their ghost numbers. Of interest to us are the ghost
number zero, one and two operators given respectively by  $\Osn(z)\bOsn(\bz)$,
$Y^+_{s+1,p}(z)\bOsn(\bz)$ as well as $a(z,\bz)\Osn(z)\bOsn(\bz)$ and
$Y^+_{s,p}(z)\bar Y^+_{s,p'}(\bz)$.  The complex conjugates of these operators
should also be added to the list. In each case the subscript $s$
characterises the $SU(2)\times SU(2)$ content of each  object. For a given
integer or $1/2$ integer $s$ the indices $p$ and $p'$ range from $-s$ to
$+s$\refs\witten .

Now consider a \kk\ tower originating from a $q$-form field in $T^{11}$. For
a given $U(1)$ charge $s$ we consider the modes at the bottom of each tower
( those which presumably have rational conformal
dimensions in the boundary gauge theory). Our observation is that these
\kk\ modes are in correspondence with the ghost number $q$ cohomology
classes in the $c=1$ theory. In Table 1 we have listed all the \kk\ modes and
the corresponding objects in the $c=1$ model. Note that the modes originating
from the components of the metric in $T^{11}$ ( which are not $q$-forms in
the internal space!) do not seem to have a counterpart in the $c=1$ side.
Furthermore the modes corresponding to an operator containing $a(z,\bar z)$
in the $c=1$ side can actually be gauged away in the \kk\ side.

The $c=1$ theory has an infinite dimensional algebra given in terms
of the volume preserving diffeomorphisms of the quadric cone
$a_1a_2 -a_3a_4=0$. It is known that the base of this cone is
isomorphic to $T^{11}$. In the context of present discussion the
$3$ complex dimensional Ricci flat cone is in fact identical to the subspace
transverse to the $D_3$-brane solution of the type IIB supergravity. Our
\kk\ background is a near horizon approximation  to this $D_3$ brane
geometry. Thus the cone seems to be the common geometrical entity in the two
very different looking theories\foot{Similar remarks can be made about
the manifolds $T^{nn}$. In this case we should consider the $c=1$ string
theory at $n$ times the self-dual radius \refs\gjm.}.

Let us consider the quadric cone $a_1a_2 -a_3a_4=0$.
This cone is singular at its apex $a_1=a_2 =a_3=a_4=0$. One can resolve the
singularity by deforming the defining equation into $a_1a_2 -a_3a_4=\mu$. From
the point of $c=1$ theory $\mu$ corresponds to the $2$-dimensional cosmological
constant. One can also consider a topological $\sigma$ model targeted on a CY
three fold near a conical singularity, for which the local equation is the
same as our quadratic expression. For both of these theories the free energies
can be evaluated as a function of $\mu$ and
can be expressed as a  genus expansion. In \refs\gv\  Ghoshal and Vafa
argued that in fact the two theories must be the same. They observed that, at
the self dual radius, the $g=0$, $1$ and $2$ contributions
to the free energy of the $c=1$ theory agree with the corresponding terms
of the free energy of the topological sigma model near the conifold
singularity. Subsequently, assuming the type II-Heterotic duality, the results
of\refs\ag\  gave further support to the Ghoshal Vafa conjecture. These
authors calculated the coefficient of the term $R^2 F^{2g-2}$ in the effective
action of the Heterotic theory compactified on $K_3\times T^2$
 and realised that for any $g$ the coefficient is also given by the genus
$g$ term of the partition function of the $c=1$ theory at the self dual
radius. More recently Gopakumar and Vafa \refs\gvafa\ have calculated the
$\sigma$ model partition function near a conifold singularity
and have proven the conjecture made in \refs\gv\ .

A better understanding of the correspondences noted above
may require the unraveling of the relevance
of the volume preserving diffeomorphisms of the cone in the $D_3$ brane
context. On the basis of the observations made in this note we would like
to think that the $c=1$ theory at the self dual radius has a role to play in
organising the chiral primaries of the boundary $SU(N)\times SU(N)$
superconformal gauge theory.

{\bf Acknowledgments:} We would like to thank Mathias Blau, Edi Gava, D.
Ghoshal, S. F. Hassan, M. Moriconi, K.S. Narain, Ashoke Sen, George Thompson
and specially  Cumrun Vafa for useful conversations.  One of us (D.P.J.) would
like to thank Abdus Salam ICTP for warm hospitality.
\vfill
\eject
Table 1: The quantum numbers of the KK spectrum of type IIB string theory on 
\ads $\times T^{nn'}$. Correspondence with the $c=1$ model at the self dual
radius holds for n=n'=1 and for each KK tower only for the modes at the bottom
of the tower. The rank of the differential form on $T^{11}$ is mapped to the 
ghost number on the $c=1$ side. The $c=1$ spectrum given in the table is at 
self dual radius and for $s\ge 0$ only. Similar analysis can be done for 
negative $s$
\bigskip
\begintable
KK excitations of IIB|Constraints on $l$ and $l'$|$c=1$ Analog\elt
$h_{\mu\nu},\, A_{\mu\nu\rho\sigma},\, B_{\mu\nu},\, \Phi$|
$l\ge \mid sn\mid $, $l'\ge \mid sn'\mid $|$\Osn(z)\bOsn(\bz)$\elt
$h_{\mu +},\, B_{\mu +},\, A_{\mu\nu\rho +},$|
$l\ge \mid sn+ 1\mid$, $l'\ge \mid sn'\mid $|
$Y^+_{s+1,p}(z)\bOsn(\bz)$\elt
$A_{\mu\nu\rho -'},\, h_{\mu -'},\, B_{\mu -'}$|
$l\ge \mid sn\mid$, $l'\ge \mid sn'-1\mid $|
$Y^+_{s,p}(z)\bOsmn(\bz)$\elt
$h_{\mu -},\, B_{\mu -},\, A_{\mu\nu\rho -}$|
$l\ge \mid sn-1\mid $, $l'\ge \mid sn'\mid$|
$\Osmn(z)\bar Y^+_{s,p'}(\bz)$\elt
$A_{\mu\nu\rho +'},\, h_{\mu +'},\, B_{\mu +'}$|
$l\ge \mid sn\mid $, $l'\ge \mid sn'+ 1\mid$|
$\Osn(z)\bar Y^+_{s+1,p'}(\bz)$\elt
$h_{\mu 5},\, A_{\mu\nu\rho 5},\, B_{\mu 5}$|$l\ge \mid sn\mid $, $l'\ge
\mid sn'\mid $|
$a(z,\bz)\Osn(z)\bOsn(\bz)$\elt
$h_{\pm\pm}$|$l\ge \mid sn\pm 2, 0\mid$, $l'\ge \mid sn'\mid $|\elt
$h_{\pm(\pm)'}$|$l\ge \mid sn\pm 1\mid$, $l'\ge \mid sn'\pm 1\mid$|
\elt
$h_{(\pm)'(\pm)'}$|$l\ge \mid sn\mid $, $l'\ge \mid sn'\pm 2, 0\mid$|
\elt
$h_{\pm 5}$|$l\ge \mid sn\pm 1\mid$, $l'\ge \mid sn'\mid $|\elt
$h_{(\pm)' 5}$ | $l\ge \mid sn\mid $, $l'\ge \mid sn'\pm 1\mid$|\elt
$h_{55}$|$l\ge \mid sn\mid $, $l'\ge \mid sn'\mid $|\elt
\vbox{\hbox{$A_{\mu\nu +-},\, A_{\mu\nu +'-'},$}\hbox{}
\hbox{$ B_{+-},\, B_{+'-'}$}}|\vbox{\hbox{}
\hbox{$l\ge \mid sn\mid $, $l'\ge \mid sn'\mid $}\hbox{}}|\vbox{\hbox{}
\hbox{$Y^+_{s,p}(z)\bar Y^+_{s,p'}(\bz)$}\hbox{}}\elt
$A_{\mu\nu\pm(\pm)'},\, B_{\pm(\pm)'}$|$l\ge \mid sn\pm 1\mid$, $l'\ge
\mid sn\pm 1\mid$|$Y^+_{s\pm 1,p}(z)\bar Y^+_{s\pm 1,p'}(\bz)$\elt
$A_{\mu\nu +5},\, B_{+5}$|
$l\ge \mid sn+1\mid$, $l'\ge \mid sn'\mid $
|$a(z,\bz)Y^+_{s+1,p}(z)\bOsn(\bz)$\elt
$A_{\mu\nu -'5},\, B_{-'5}$|
$l\ge \mid sn\mid$, $l'\ge \mid sn'-1\mid $
|$a(z,\bz)Y^+_{s,p}(z)\bOsmn(\bz)$\elt
$A_{\mu\nu -5},\, B_{-5}$|
$l\ge \mid sn-1\mid$, $l'\ge \mid sn'\mid $
|$a(z,\bz)\Osmn(z)\bar Y^+_{s,p'}(\bz)$\elt
$A_{\mu\nu +'5},\, B_{+'5}$|
$l\ge \mid sn\mid$, $l'\ge \mid sn'+1\mid $
|$a(z,\bz)\Osn(z)\bar Y^+_{s+1,p'}(\bz)$
\endtable
\footatend\vfill\supereject\immediate\closeout\rfile\writestoppt
\baselineskip=14pt\centerline{{\bf References}}\bigskip{\frenchspacing%
\parindent=20pt\escapechar=` \input refs.tmp\vfill\eject}\nonfrenchspacing
\bye